\documentclass{aastex701}

\begin{document}

\title{The Solar System Notification Alert Processing System (SNAPS): Public access to SNAPS data and products}

\author[orcid=0000-0003-4580-3790]{David E. Trilling}
\affiliation{Department of Astronomy and Planetary Science, Northern Arizona University}
\affiliation{School of Informatics, Computing, and Cyber Systems, Northern Arizona University}
\email[show]{david.trilling@nau.edu}

\author[orcid=0000-0002-0826-6204]{Michael Gowanlock}
\affiliation{School of Informatics, Computing, and Cyber Systems, Northern Arizona University}
\affiliation{Department of Astronomy and Planetary Science, Northern Arizona University}
\email{michael.gowanlock@nau.edu}

\author[orcid=0009-0006-2731-5522]{Revanth Munugala}
\affiliation{School of Informatics, Computing, and Cyber Systems, Northern Arizona University}
\affiliation{Department of Astronomy and Planetary Science, Northern Arizona University}
\email{rm2878@nau.edu}

\author[orcid=0000-0002-6676-1713]{Daniel R. Kramer}
\affiliation{School of Informatics, Computing, and Cyber Systems, Northern Arizona University}
\affiliation{Department of Astronomy and Planetary Science, Northern Arizona University}
\email{drkspace@gmail.com}

\author[orcid=0000-0002-6292-9056]{Maria Chernyavskaya}
\affiliation{Department of Astronomy and Planetary Science, Northern Arizona University}
\email{mc3944@nau.edu}

\author{Erin Clark}
\affiliation{Department of Astronomy and Planetary Science, Northern Arizona University}
\email{emc635@nau.edu}

\author{Graceson Mule}
\affiliation{School of Informatics, Computing, and Cyber Systems, Northern Arizona University}
\email{gjm256@nau.edu}

\author{Savannah Chappus}
\affiliation{School of Informatics, Computing, and Cyber Systems, Northern Arizona University}
\email{sjc497@nau.edu}

\begin{abstract}
The Solar System Notification Alert Processing System, SNAPS, 
is a downstream broker that ingests moving object data from ZTF and LSST and serves these data and derived properties to the public.
This document describes how users can access our SNAPS data and products. This is intended to be a living document that will be updated on the arXiv when significant improvements are made to our data access schemes, and will therefore always contain the most up to date information about interacting with our databases and infrastructure. This is version~1.0.
\end{abstract}



\section{Introduction}

Astronomy is now a ``big data'' science, with peta-scale surveys collecting all-sky data that a large user base can access for a wide range of investigations. The arrival of the Legacy Survey of Space and Time (LSST), which will be carried out with the Vera C. Rubin Observatory in Chile, will be the largest-yet survey.
This project will observe 
some 20,000~deg$^2$ many times over ten years and will produce a catalog of some 40~billion astronomical sources, including 5~million Solar System objects and around 500~million moving object observations
\citep{lsstbook,ivezic2019}.
To carry out science with such large-scale surveys, comprehensive software infrastructure is required, and the LSST project has championed the concept of ``brokers,'' which ingest data from an alert stream and serve a subset of this data and derived products to users interested in specific topics.

Here we present a description of the public-facing aspects of the Solar System Notification Alert Processing System (SNAPS).
SNAPS is the only Solar System-dedicated broker in the LSST ecosystem. The architecture of SNAPS, as well as some early science and our first data release (SNAPShot1, or SS1), are presented in 
\citet{ss1}.
In this paper we briefly describe what SNAPS does and does not do (Section~\ref{summary}), and then present details of our web service and interface (Section~\ref{access}). In Section~\ref{future} we briefly describe the timelines and goals for future science and technical steps.

The first version of this document will be submitted to the arXiv in April, 2026. 
Subsequent significant updates to SNAPS infrastructure will be described in future versions of this paper, which we will place on the arXiv to supersede earlier versions. Thus, the arXiv link and DOI associated with this document are intended to be the permanent reference to SNAPS, and SNAPS users are requested to cite either or both this paper and our SNAPS architecture paper 
\citep{ss1}. This document is paper version~1.0.

\section{Brief summary of SNAPS \label{summary}}

SNAPS is purpose-built to provide a number of products and access to the potential user community, but, naturally, it cannot provide all things to all people. Here we briefly describe what SNAPS is and does, and what it does not do.

SNAPS generally operates as a ``downstream'' broker. This means that rather than listening to the complete outgoing alert stream from LSST or similar surveys \citep[such as the Zwicky Transient Facility (ZTF),][]{ztf}, we instead listen to a re-broadcast of Solar System-only alerts that is distributed by the ANTARES broker 
\citep{antares}. Therefore, SNAPS does not handle any sources that are not already designated as known moving objects (generally Solar System objects, but interstellar objects can also be present in the stream).

The core SNAPS processing tasks are the following. (1) SNAPS ingests data --- typically as alerts --- from all-sky surveys such as ZTF and LSST. We store observational data (RA, Dec, time, magnitude), meta-data (elongation of the source, FWHM, etc.), and postage stamps (science image, template image, and difference image) for each observation. A user can retrieve these values for any individual measurement or set of measurements (e.g., all measurements for a given object). We ingest data for known asteroids and known comets (and interstellar objects). Data from each survey is separately stored in our database --- there are separate ZTF and LSST tables/collections --- though future work includes retrieving merged data records (example: a query for all data on a given object, from any survey).
(2) SNAPS detects individual outliers, which are objects whose attributes have changed since its last observation. The most obvious example here is an asteroid that shows new signs of activity. The details of these detection algorithms will be presented in subsequent papers. (3) SNAPS detects population outliers, which are objects whose properties are intrinsically unusual. Several algorithms can be used to detect these outliers, as summarized in \citet{outliers}.
(4) SNAPS serves these results to interested users in various ways, as described in this paper.

What SNAPS does not do includes the following. (1) We do not carry out orbit linking to
match new observations to previously known objects. We do not ingest any observations that are not associated with known objects. 
(Note that this limitation means that the first few observations of newly discovered objects likely will not be present in our database, though in principle these could be recovered after the fact.)
We do not discover Solar System bodies.
(2) We do not look for orbit changes of any kind.
(3) We do not ingest or serve any data related to non-moving objects. 

\section{Public access to SNAPS \label{access}}

The two primary methods of interacting with our SNAPS database are through our web page and through our API.
Generally, web and API access divide the use cases between ``browsing'' or investigating an individual object of interest, through our web page, or ``bulk analysis'' --- for example, to study a group of objects --- through the API. 
At the moment, our web page allows ``browse''-level access, as is described here. In the near future, when our API is functional and public-facing, an updated version of this paper will be published in which that ``bulk''-level analysis is described.

The SNAPS webpage includes a user-friendly interface to browse our database catalog, observation data, and view postage stamps. This web portal can be found at \url{https://snaps.nau.edu/} . The webpage is supported by two backend databases: (1) An Oracle database to store user-specific structured data such as recent queries, favorites, among others. (2) A Mongo database that supports querying our ingested alerts from various catalogs, our data releases, and derived properties, among others records.

The core functionality of our webpage is described as follows.

\begin{itemize}
\item Users can create a free account, which enables storing ``Favorites'' and recent queries. Users who are not logged in are somewhat rate-limited to mitigate denial-of-service attacks and other malicious interactions.
Users who are logged in have a more generous access privileges, but some protections still exist; ``power users'' can contact us to discuss further relief of these limitations.
\item Individual objects can be queried based on their number (example: asteroid 12345), their provisional designation
(example: 1995 UW49), or their packed designation (example: 
J95U49W). The search bar has string completion to suggest objects in our database that begin with the input characters.
\item At present, users can query either the ZTF catalog or the LSST catalog. At present, the ZTF catalog is divided into numbered objects (collections referred to as SNAPShot1 or SNAPShot2) or ``provisional'' objects (those objects that have not yet been assigned a number).
(Future work includes a ``joint query'' where observations and properties for a given asteroid that includes data from any/all surveys in our catalogs will be reported.)
\item The search query returns all observations from that collection for that object: 
a unique identifier (ZTFid, ssObjectId if from LSST, etc.), time, magnitude, error, filter, and geometric values (example: heliocentric and geocentric distances and phase angle).
For each observation, H (Solar System absolute magnitude) in the observed band is calculated using the HG system \citep{bowell1989}.
We also report orbital elements and other data from the Minor Planet Center (MPC).

\item In addition to this, 
for objects with more than 50~observations
we also show derived data products
(example: phase slope parameter G, lightcurve period and amplitude, and derived color in all relevant filter combinations). 
\item Our webpage also includes a robust plot functionality to view observational records (absolute magnitude as a function of time), phased lightcurve (absolute magnitude as a function of phase), and phase curve (reduced magnitude as a function of phase angle) plots.
The plot windows are interactive (through the {\tt plotly} technology), allowing zooming, downloading the plot as PNG, selecting/unselecting observations in any reported filter, and so on.

\item Our webpage also supports on-the-fly period finding using the Lomb-Scargle algorithm~\citep{lomb1976least,scargle1982studies} across a user-selected range of periods and date ranges. The result is a periodogram. Users can rephase the observed data to a peak period solution from the periodogram, or any other custom period. This task runs in the browser.
\item For users who are logged in, all recent queries are stored in our database and shown on the homepage 
(for example, to run the same query at a future time).
\item We also include a functionality for users to ``favorite'' an object and add notes regarding the observations.
\item We allow users to view the postage stamps for all observations with time stamps. We use JS9 to support showing the postage stamps in FITS format and allow downloads as well. Some in-browser interactivity is allowed through the JS9 interface. For all observations we serve three postage stamps: science, template, and difference images.
\item Lastly, our webpage lists 
some our publications and includes funding acknowledgments.
We also have a contact form for users to communicate suggestions, queries, or requests, and to report any issues.
\end{itemize}

As of this writing, our comet
functionality is not fully mature. Users can query comets in the ``Provisional/ZTF'' query (LSST is not yet serving comet data), 
and magnitude as a function of time will be reported,
but 
as of this writing no comets have more than 50~observations, so we do not serve any derived properties.


\section{Future work \label{future}}

There are two key feature upgrades that are coming in the near future.
(1) We have an internal-only API that has limited functionality; in the near future we will make this public-facing so that external users can interact with our databases in ``bulk query'' mode. This will address several of the limitations described above. This paper will be updated when this external-facing API is functioning. 
(2) We will start to publish an outgoing Kafka stream that reports unusual objects. There are two categories of unusual, or outlier, asteroids: individual outliers, whose properties have changed compared historical observations (example: asteroids that start to show activity), and population outliers, which are asteroids whose properties are intrinsically different from those of other asteroids. The methodology for our population outlier detection pipeline is described in \citet{outliers}. We are currently developing the algorithmic methodology for detecting individual outliers.
This paper will be updated when the Kafka stream is live.

Alongside those future upgrades we will make several updates to the webpage as follows: (1) Allowing queries based on object properties rather than solely querying based on object identifiers. (2) Posting individual and population outlier rankings. Furthermore, in the future we will (3) implement cross-catalog queries.

For the foreseeable future, ZTF will continue to operate an all-sky survey that will produce alerts. We will continue to ingest alerts from ZTF into our database. 
The LSST survey will begin in 2026, and alert stream production will probably ramp up as more and better template images become available. Within the first year of full-scale LSST operations, several million asteroids will be observed \citep{kurlander}, almost all of which will be new discoveries.

Our team is working on a number of science and computer science projects that will be published in the near future. These include identifying a number of asteroids and asteroid populations of interest and the implementation of new algorithms to define the fidelity of asteroid period solutions.

\begin{acknowledgments}

We acknowledge support from NASA award
80NSSC25K7728,
NSF awards 2042155 and 2206796, and DOE/LLNL award B657367.
This work is supported by the Arizona Board of Regents and the Technology Research Initiative Research Fund (TRIF) Small Research Equipment Acquisition Program (SREAP).
We thank NAU's Chris Coffey for his exemplary leadership of our High Performance Computing facility. We thank Andrew McNeill (BGSU) for many useful discussions about the direction and utility of SNAPS.

\end{acknowledgments}

%
\facilities{ZTF, LSST}

\software{astropy~\citep{2013A&A...558A..33A,2018AJ....156..123A,2022ApJ...935..167A}, LSP-GPU~\citep{gowanlock2021fast}, plotly~\citep{plotly}, SciPy~\citep{2020SciPy-NMeth}}



\bibliography{snapsaccess}{}
\bibliographystyle{aasjournalv7}



\end{document}